\documentclass[aps,nofootinbib,preprint,showpacs,superscriptaddress]{revtex4}
\date{\today}
\usepackage{graphicx}

\begin{document}

\title{Microscopic mechanism of charged-particle radioactivity and generalization of the
Geiger-Nuttall law}

\author{C. Qi}
\affiliation{School of Physics, and State Key Laboratory of Nuclear
Physics and Technology, Peking University, Beijing 100871, China}
\affiliation{KTH (Royal Institute of Technology), Alba Nova
University Center, SE-10691 Stockholm, Sweden}

\author{F.R. Xu}
\affiliation{School of Physics, and State Key Laboratory of Nuclear
Physics and Technology, Peking University, Beijing 100871, China}
\affiliation{Center for Theoretical Nuclear Physics, National
Laboratory for Heavy Ion Physics, Lanzhou 730000, China}

\author{R.J. Liotta}
\affiliation{KTH (Royal Institute of Technology), Alba Nova
University Center, SE-10691 Stockholm, Sweden}

\author{R. Wyss}
\affiliation{KTH (Royal Institute of Technology), Alba Nova
University Center, SE-10691 Stockholm, Sweden}

\author{M.Y.~Zhang}
\affiliation{School of Physics, and State Key Laboratory of Nuclear
Physics and Technology, Peking University, Beijing 100871, China}

\author{C. Asawatangtrakuldee}
\affiliation{School of Physics, and State Key Laboratory of Nuclear
Physics and Technology, Peking University, Beijing 100871, China}

\author{D. Hu}
\affiliation{School of Physics, and State Key Laboratory of Nuclear
Physics and Technology, Peking University, Beijing 100871, China}

\begin{abstract}
A linear relation for charged-particle emissions is presented
starting from the microscopic mechanism of the radioactive decay. It
relates the logarithms of the decay half-lives with two variables,
called $\chi'$ and $\rho'$, which depend upon the $Q$-values of the
outgoing clusters as well as the masses and charges of the nuclei
involved in the decay. This relation explains well all known cluster
decays. It is found to be a generalization of the Geiger-Nuttall law
in $\alpha$ radioactivity and therefore we call it the universal
decay law. Predictions on the most likely emissions of various
clusters are presented by applying the law over the whole nuclear
chart. It is seen that the decays of heavier clusters with non-equal
proton and neutron numbers are mostly located in the trans-lead
region. The emissions of clusters with equal protons and neutrons,
like $^{12}$C and $^{16}$O, are possible in some neutron-deficient
nuclei with $Z\geq54$.
\end{abstract}

\pacs{21.10.Tg, 23.60.+e, 23.70.+j, 27.60.+j, 27.90.+b}

\maketitle

\section{Introduction}

Charged-particle emissions are among the most important decay modes
of atomic nuclei. Almost all observed proton-rich exotic nuclei
starting from $A \sim 150$ are $\alpha$ radioactive~\cite{Audi03}. A
substantial number of proton decays have been observed in
proton-drip-line nuclei around the rare earth region~\cite{Son02}.
The spontaneous emission of charged fragments heavier than the
$\alpha$ particle (cluster decay) was predicted in
Ref.~\cite{Sandulescu80} and later established experimentally in
trans-lead mother nuclei decaying into daughters around the doubly
magic nucleus $^{208}$Pb~\cite{Rose84,Hou89,Price89,Bonetti07}. Even
a second island of cluster radioactivity was predicted in trans-tin
nuclei decaying into daughters close to the doubly magic nucleus
$^{100}$Sn~\cite{Grei89}.

A number of theoretical models were proposed to describe the
charged-particle decay
process~\cite{Poe91,Shi85,Lovas98,Delion94,Dumi94,Buck89,Buck93,Blen88,Delion06a,Mang64,Arima74}
(see also Refs.~\cite{xu06,zhang08,Mad08,Routray09,Bhagwat08,Ren09}
for very recent calculations). In general the decay process, ranging
from proton to heavier cluster radioactive decays, can be described
by a two-step mechanism~\cite{Tho54}. The first step refers to the
formation of the particle and its motion on the daughter nuclear
surface. In the second step the cluster, with the formation
amplitude and corresponding wave function thus determined, is
assumed to penetrate through the centrifugal and Coulomb
barriers~\cite{Gam28,Liotta86,Poe02,Blen91}. This second step is
well understood since the pioneering work of Gamow~\cite{Gam28}. In
macroscopic models, cluster decay is treated as the quantum
tunneling process of an already preformed
particle~\cite{Buck89,Buck93,xu06,zhang08,Routray09,Bhagwat08,Ren09},
where features like the probability that the cluster is formed on
the nuclear surface are ignored. In these models the clusterization
process is included in an effective fashion by introducing
quantities adjusted to reproduce as many measured half-lives as
possible. Such semiclassical models are being successfully applied
even at present, although in some cases microscopical ingredients
are also included~\cite{Buck89}. In microscopic theories the
formation amplitude is evaluated starting from the single-particle
degrees of freedom of the neutrons and protons that eventually
become the cluster. This is generally a formidable task which
requires advanced computing facilities as well as suitable
theoretical schemes to describe the clusterization
process~\cite{Delion94,Lovas98,Mang64,Arima74}.

On the other hand, this variety of theoretical models may serve as a
guide to our searching of semiclassical relations in the radioactive
decay. The first striking correlation in $\alpha$ decay systematics
was noted by Geiger and Nuttall~\cite{gn}. This relates the decay
half-lives $T_{1/2}$ and decay energies $Q_{\alpha}$ as,
\begin{equation}\label{gn-o}
\log T_{1/2}=aQ_{\alpha}^{-1/2}+b,
\end{equation}
where $a$ and $b$ are constants. Nowadays it is understood that the
$Q$-value dependence in Eq.~(\ref{gn-o}) is a manifestation of the
quantum penetration of the $\alpha$-cluster through the Coulomb
barrier (see, for example, Ref.~\cite{Buck90}). But this equation
ignores the probability that the $\alpha$-particle is formed on the
nuclear surface starting from its four constituent nucleons moving
inside the mother nucleus. The linear relation~(\ref{gn-o}) has been
found to hold well for the ground-state to ground-state decays of
even-even nuclei in the same major shell with fixed proton number.
However, the Geiger-Nuttall law in the form of Eq.~(\ref{gn-o}) has
limited prediction power since the coefficients $a$ and $b$ change
for the decays of different isotopic series~\cite{vs}. Intensive
works have been done trying to generalize the Geiger-Nuttall law for
a universal description of all observed $\alpha$ decay
events~\cite{vs,Brown92,Moll97,Royer00,Poe06,Denisov09}. For
example, in the work of Viola and Seaborg~\cite{vs}, the $a$ and $b$
coefficients of Eq.~(\ref{gn-o}) are assumed to be linearly
dependent upon the charge number of the daughter nucleus. But the
physical origin of this dependence is not clear. Empirical linear
relations were also found in the cases of proton
decays~\cite{Delion06,Mede07,Dong09}, heavier cluster
decays~\cite{Ren04,Poe02,Horoi04,Bal04} and both $\alpha$ and
heavier cluster decays~\cite{Ren08,San08,Delion09}. Reviews on
existing empirical relations can be found in
Refs.~\cite{Poe06,San08} and thus will not be detailed here. In
particular, some recent searches of correlations in radioactive
decays start from the macroscopic description of the decay process
with a concentration on the Coulomb barrier
penetrability~\cite{Buck90,Poe06,Brown92,Ren08}, which are
physically more sound than mere empirical relations. But in these
macroscopic approaches one has to assume an effective interaction
between the cluster and the core~\cite{Brown92,Buck90}. Besides, an
effective spectroscopic factor (formation amplitude) has to be
introduced~\cite{Poe06} which, however, is model-dependent and
sensitive to details of the effective interaction.

In a recent Letter~\cite{qi08}, we have introduced a linear
universal decay law (UDL) starting from the microscopic mechanism of
the charged-particle emission. Our aim was to find a general
framework valid for all clusters, which may be used in the future as
a gauge to probe effective formulas. This is an interesting subject
in itself, but perhaps even more important is that it may help in
the ongoing searching of new cluster decay modes from superheavy
nuclei~\cite{Poe02}. The UDL relates the half-lives of monopole
radioactive decays with the $Q$-values of the outgoing particles as
well as the masses and charges of the nuclei involved in the decay,
reflecting quite well the systematical trend of experimental data.
In this paper we will complete the brief presentation given in
Ref.~\cite{qi08} with details of the construction of the formula and
approximations leading to it. Besides, we present the predictions of
the UDL on the most likely emissions of various clusters.

In Section II is the Formalism. In Section III systematics of
experimental $\alpha$ and cluster decay half-lives are analyzed and
compared with the corresponding calculations. In Section IV possible
observations of new cluster decays are suggested. A Summary and the
Conclusions are in Section V.

\section{Formalism}

In a classic paper~\cite{Tho54}, Thomas derived the expression of
the cluster decay width by evaluating the residues of the
corresponding S-matrix in the framework of the R-matrix
theory~\cite{Lan58}. The decay half-life thus obtained has the form,
\begin{equation}\label{life}
T_{1/2}=\frac{\hbar\ln2}{\Gamma_c}=\frac{\ln2}{\nu} \left|
\frac{H_l^+(\chi,\rho)}{RF_c(R)} \right|^2,
\end{equation}
where $\Gamma_c$ is the decay width, $\nu$ the outgoing velocity of
the charged particle carrying an angular momentum $l$. $R$ is the
distance between the corresponding centers of mass of the cluster
and daughter nucleus, which should be large enough that the nuclear
interaction is negligible. $H^+_l$ is the Coulomb-Hankel function
and its arguments are standard, i.e., $\rho=\mu\nu R/\hbar$ and the
Coulomb parameter is $\chi = 2Z_cZ_de^2/\hbar\nu$ with $\mu$ being
the reduced mass of the cluster-daughter system and $Z_c$ and $Z_d$
the charge numbers of the cluster and daughter nucleus,
respectively. Eq.~(\ref{life}) contains the two-step mechanism
mentioned above. The quantity $F_c(R)$ is the formation amplitude of
the decaying particle at distance $R$, which is usually evaluated as
the overlap between the mother wave function and the antisymmetrized
tensor product of the daughter and cluster wave functions. The
penetrability is proportional to $|H_l^+(\chi,\rho)|^{-2}$. This
equation is the basis of all microscopical calculations of
radioactive decay processes~\cite{Lovas98,Delion06}. It is valid for
all clusters and for spherical as well as  deformed cases. The ratio
$N_l =RF_c(R)/H_l^+(R)$, and therefore the half-life itself, is
independent of the radius $R$~\cite{Lovas98}. In
Ref.~\cite{Maglione08} it is shown that the expression of
Eq.~(\ref{life}) coincide with the quantum-mechanical interpretation
of the half-life as the outgoing flux per unit of time.

In what follows we will apply the exact expression of
Eq.~(\ref{life}). Our aim is to find a few quantities that determine
the half-life. Expanding in these quantities we hope to be able to
find, at the lowest order of perturbation, an expression of the
half-life which is as simple as the Geiger-Nuttall law but valid in
general, i.e., for all isotopic series as well as all type of
clusters. This is possible since Eq.~(\ref{life}) itself is valid in
general. The number of variables that we have to look for should be
small for cases of interest, i.e., for the decay of  medium and
heavier nuclei. In fact most interesting is the predicting power
with respect to superheavy nuclei, which are at the center of
attention of present experimental activities. With this in mind we
notice that the Coulomb-Hankel function can be well approximated by
an analytic formula, which for the $l=0$ channel reads~\cite{fro57},
\begin{equation}
H^+_0(\chi,\rho) \approx (\cot
\beta)^{1/2}\exp\left[\chi(\beta-\sin\beta\cos\beta)\right],
\end{equation}
where the cluster $Q$-value is $Q_c=\mu \nu^2/2$ and
\begin{equation}\label{cosb}
\cos^2\beta=\frac{\rho}{\chi} = \frac{Q_c R}{e^2Z_cZ_d}.
\end{equation}

One sees that $\cos^2\beta$ would be a small quantity if $Z_cZ_d$ is
large. In this case one can expand the last term in a power series
of $\cos\beta$ (with $\beta= \arccos(\cos\beta))$ as,
\begin{equation}
\beta-\sin\beta\cos\beta=\frac{\pi}{2}-2\cos\beta +
\frac{\cos^3\beta}{3} + \frac{\cos^5\beta}{20}+\cdots,
\end{equation}
and for medium and heavier nuclei (the heavier the better) terms
beyond the third order can be neglected. One obtains,
\begin{equation}
\log \frac{|H^+_0(\chi,\rho)|^2}{\cot\beta} \approx
\frac{2\chi}{\ln 10} \left[\frac{\pi}{2}-
2\left(\frac{\rho}{\chi}\right)^{1/2} + \frac{1}{3}
\left(\frac{\rho}{\chi}\right)^{3/2} \right],
\end{equation}
and therefore,
\begin{eqnarray}
\nonumber \log T_{1/2} &\approx&\frac{2\chi}{\ln 10}
\left[\frac{\pi}{2}- 2\left(\frac{\rho}{\chi}\right)^{1/2} +
\frac{1}{3} \left(\frac{\rho}{\chi}\right)^{3/2} \right] \\
\nonumber&&+ \log \left(\frac{\cot\beta\ln 2}{\nu
R^2|F_c(R)|^2}\right), \\
\end{eqnarray}
which is dominated by the first two terms. For the radius $R$ in
above equation, one can take the standard value of
$R=R_0(A_d^{1/3}+A_c^{1/3})$ with $R_0\sim
1.2$~fm~\cite{Delion06,qi08}. Defining the factors $\chi'$ and
$\rho'$ as,
\begin{eqnarray}\label{chirhop}
\chi' &=& \frac{\hbar}{e^2\sqrt{2m}}\chi=Z_cZ_d\sqrt{\frac{A}{Q_c}},
\nonumber
\\
\nonumber \rho' &=& \frac{\hbar}{\sqrt{2mR_0e^2}} \left( \rho
\chi\right)^{1/2}
\\
&=& \sqrt{AZ_{c} Z_d(A_d^{1/3}+A_c^{1/3})},
\end{eqnarray}
where $A=A_d A_c/(A_d+A_c)=\mu/m$ and $m$ is the nucleon mass
(within the errors of our treatment we take $mc^2 \approx 938.9$~MeV
and $\hbar c=197.3$~MeV~fm), one gets
\begin{eqnarray}\label{gn-1}
\nonumber\log T_{1/2} &=& \frac{\sqrt{2M}e^2\pi}{\hbar\ln10}\chi' -
\frac{4e\sqrt{2MR_0}}{\hbar\ln10}\rho'\\
\nonumber&&+ \log \left(\frac{\cot\beta\ln 2}{\nu
R^2|F_c(R)|^2}\right)+ o(3), \\
 &=&a\chi' + b\rho' - 2\log|F_c(R)| + c,
\end{eqnarray}
where $a$ and $b$ are constants and $o(3)$ corresponds to the
remaining small terms in the Coulomb penetration. The terms $o(3)$
and $\log \cot\beta/(\nu R^2)$ change rather smoothly for the decay
cases of interest and may be safely approximated as a constant $c$.

A straightforward conclusion from Eq.~(\ref{gn-1}) is that $\log
T_{1/2}$ depends linearly upon $\chi'$ and $\rho'$. Still the strong
dependence of the formation probability upon the cluster size has to
be taken into account by Eq.~(\ref{gn-1}). This seems to be a
difficult task, since the formation probability is strongly
dependent upon the nuclear structure of the nuclei to be analyzed.
In other words, if such a simple linear relation is correct one has
to be able to demonstrate that the formation amplitude depends only
linearly upon $\chi'$, $\rho'$ or an additional variable. We found
that this is indeed the case by exploiting the property that for a
given cluster $N_0\equiv RF_c(R)/H_0^+(\chi,\rho)$ does not depend
upon $R$. Using the approximations leading to Eq. (\ref{gn-1}) one
readily obtains the relation,
\begin{equation}
\log \left|\frac{R'F_c(R')}{RF_c(R)}\right|^{-2} \approx
\frac{4e\sqrt{2M}}{\hbar\ln10}\left(\sqrt{R_0'}-\sqrt{R_0}\right)\rho',
\end{equation}
where $R'=R'_0(A_d^{1/3}+A_c^{1/3})$ is a value of the radius that
differs from $R$. The above equation can also be written as,
\begin{equation}\label{foramp}
\log \left|RF_c(R)\right|\approx \log \left|R'F_c(R')\right|+
\frac{2e\sqrt{2M}}{\hbar\ln10}\left(\sqrt{R_0'}-\sqrt{R_0}\right)\rho'.
\end{equation}
Since for a given cluster any nuclear structure would be carried by
the terms $RF_c(R)$ and $R'F_c(R')$ in exactly the same fashion,
Eq.~(\ref{foramp}) implies that the formation amplitude is indeed
linearly dependent upon $\rho'$. Therefore one can
write~\cite{qi08},
\begin{equation}\label{gn-2}
\log T_{1/2}=a\chi' + b\rho' + c.
\end{equation}
The coefficients $b$ and $c$ in this relation are different from
that of Eq.~(\ref{gn-1}) since the terms $ b\rho'+ c$ have to
include the effects that induce the clusterization in the mother
nucleus. This relation holds for the monopole radioactive decays of
all clusters and we called it the UDL~\cite{qi08}. The relation can
be easily generalized to include the $l\neq0$ decay cases by taking
the effects of the centrifugal potential on the barrier
penetrability into account~\cite{fro57}.

It can be easily recognized that the UDL includes the Geiger-Nuttall
law as a special case since $\rho'$ remains constant for a given
$\alpha$-decay chain and $\chi'\propto Q_c^{-1/2}$. Besides, one
basic assumption behind the relation (\ref{gn-2}) is that one can
define a proper radius $R'$ that leading to a stable formation
amplitude $F_c(R')$ for all cluster radioactivities. In the next
Section we will probe these conclusions, and the approximations
leading to them.

\section{Systematics of experimental data}

In this Section we will analyze ground-state to ground-state
radioactive decays of even-even nuclei. We take all $\alpha$ decay
events from emitters with $78\leq Z\leq108$ for which experimental
data are available to us. We take the data from the latest
compilations of Refs.~\cite{Audi03,Audi03a} and the lists of
Refs.~\cite{zhang08,Pei07}. For the decay of heavier clusters we
have selected 11 measured events ranging from $^{14}$C to
$^{34}$Si~\cite{Bonetti07}. The branching ratios of the cluster
decays relative to the corresponding $\alpha$ decay are in the range
of $10^{-9}$ to $10^{-16}$. The partial half-lives of observed
cluster decays are between $10^{11}$s and $10^{28}$s.

\subsection{Experimental constraint on the formation amplitude}

\begin{figure}
\includegraphics[scale=0.6]{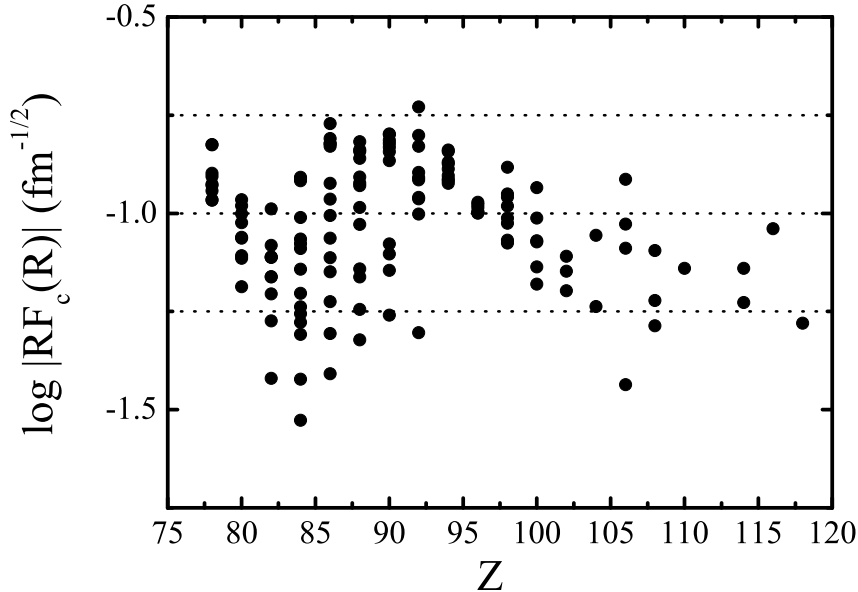}\\
\caption{The $\alpha$ decay formation amplitudes $\log |RF_c(R)|$ as
a function of the charge number of the mother nucleus
$Z$.}\label{fig1}
\end{figure}

The formation amplitude $F_c(R)$ reflects the nuclear structure
effect on the cluster decay process. According to Eq.~(\ref{life})
the formation amplitude $F_c(R)$ can be extracted from experimental
data as,
\begin{equation}
\log |RF_c(R)|=-\frac{1}{2}\log T^{{\rm Expt.}}_{1/2} +\frac{1}{2}
\log \left[ \frac{\ln 2}{\nu}|H^+_0(\chi,\rho)|^2\right].
\end{equation}
By using $R_0=1.2$~fm we evaluated the function $\log |RF_c(R)|$
corresponding to $\alpha$ clusters to obtain the results plotted in
Fig.~\ref{fig1}. One sees that as a function of the charge number of
the emitters the formation probabilities are located in the range
$\log |RF_c(R)|=-1.5\sim-0.75$~fm$^{-1/2}$ with about half of the
data below -1.0~fm$^{-1/2}$. The formation amplitude is therefore in
the range $F_c(R)=(0.03 \sim 0.18)/R$ (where $R$ is in fm and
$F_c(R)$ in fm$^{-3/2}$). One thus confirms that for a given cluster
the formation amplitude is constant within an order of magnitude.
The stability of the $\alpha$ decay formation amplitude indicates
that the linear relation described by Eq.~(\ref{gn-2}) is not as
unexpected as one might have assumed.

It is seen from Fig.~\ref{fig1} that in a few cases the formation
amplitudes become small, i.e., with the $\log
|RF_c(R)|<-1.4$~fm$^{-1/2}$. These correspond to the $\alpha$ decays
of nuclei $^{194}$Pb, $^{208,210}$Po, $^{212}$Rn and $^{266}$Sg
(note that the error in the experimental half-life of this nucleus
is still large~\cite{Audi03,Dvo06}). The $\alpha$ formation
amplitudes may have been significantly reduced in the former four
nuclei which are approaching the $Z=82$ and/or $N=126$ shell
closures.

\begin{figure}
\includegraphics[scale=0.6]{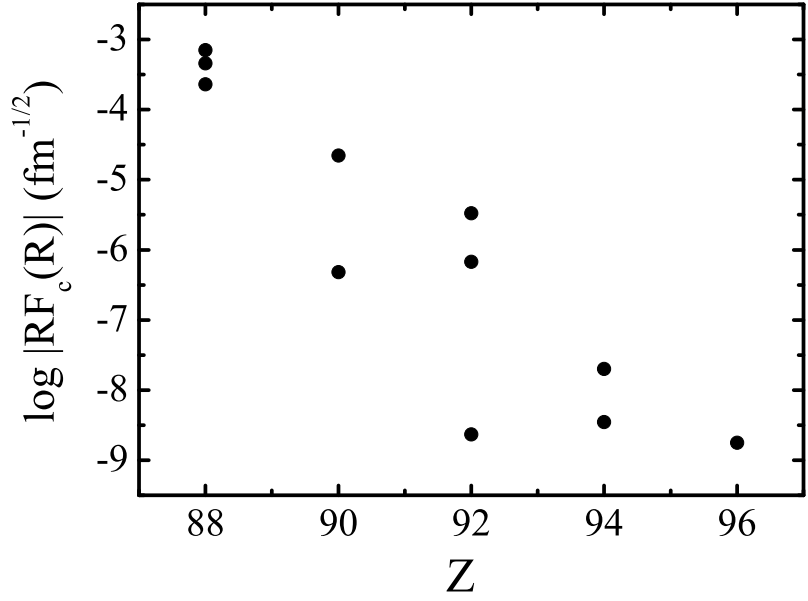}\\
\caption{Same as Fig.~\ref{fig1} but for those of heavier cluster
decays.}\label{fig2}
\end{figure}

Following the same procedure as above we evaluated $\log |RF_c(R)|$
for observed heavy clusters, as seen in Fig.~\ref{fig2}. One sees
that now $\log |RF_c(R)|$ is in the range -9 to -3, i.e.,
$F_c(R)=(10^{-9}-10^{-3})/R$ fm$^{-3/2}$. Given the variety of
clusters (from $^{14}$C to $^{34}$Si) involved in the Figure this
wide range of 6 orders of magnitude is also expected.

\begin{figure}
\includegraphics[scale=0.6]{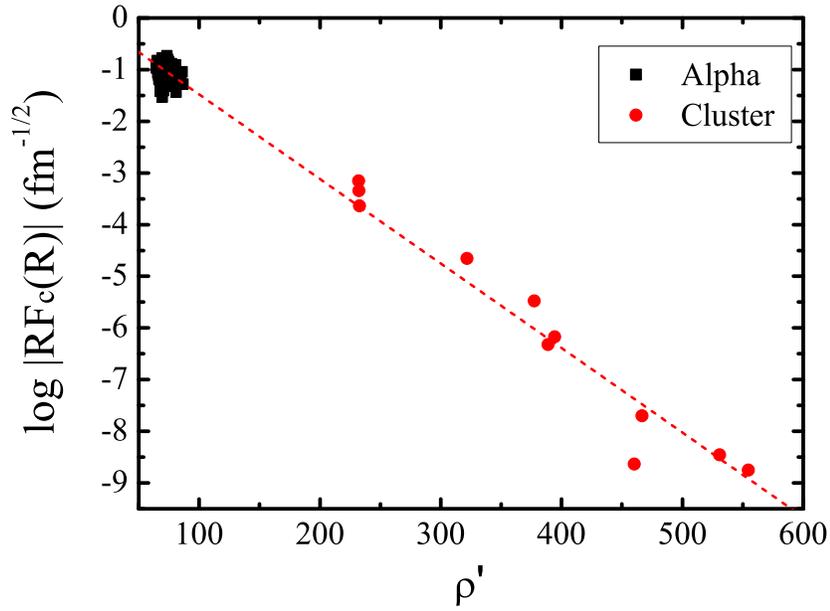}\\
\caption{(Color online) The formation amplitudes $\log |RF_c(R)|$
for both $\alpha$ and cluster decays as a function of
$\rho'$.}\label{fig3}
\end{figure}

We are now in a position to probe the validity of the linear
relation between the logarithm of the formation probability as a
function of $\rho'$ as implied by Eq.~(\ref{foramp}). As seen in
Fig.~\ref{fig3} that relation holds rather well. The majority of
available experimental data corresponds to $\alpha$-decay. Since the
formation amplitude of a given type of cluster is rather constant
one sees in the Figure an accumulation of black points around $\log
|RF(R)| \approx-1.2$~fm$^{-1/2}$. This corresponds to $\alpha$-decay
events. For the other clusters the formation probabilities decrease,
as expected. The important point that one can make from the Figure
is that the predicted linear trend is confirmed. The linear trend of
experimental data ensures that one can find an radius $R'$ where
$F_c(R')$ remain constant for all radioactivities. The $R'$ value
can be determined through a fitting procedure.

There is a deviation to this trend at $\rho'\approx 460$ which
corresponds to the decay of
$^{234}\text{U}\rightarrow~^{206}\text{Hg}+~^{28}\text{Mg}$. One may
expect that the formation of $^{26}$Ne is more favored in
$^{234}\text{U}$ (with the daughter system of $^{208}$Pb). However,
the decay emitting $^{26}$Ne is hindered due to the presence of a
much lower $Q_c$ value (i.e., larger $\chi'$).

\subsection{Systematics with the UDL}

\begin{figure}
\includegraphics[scale=0.6]{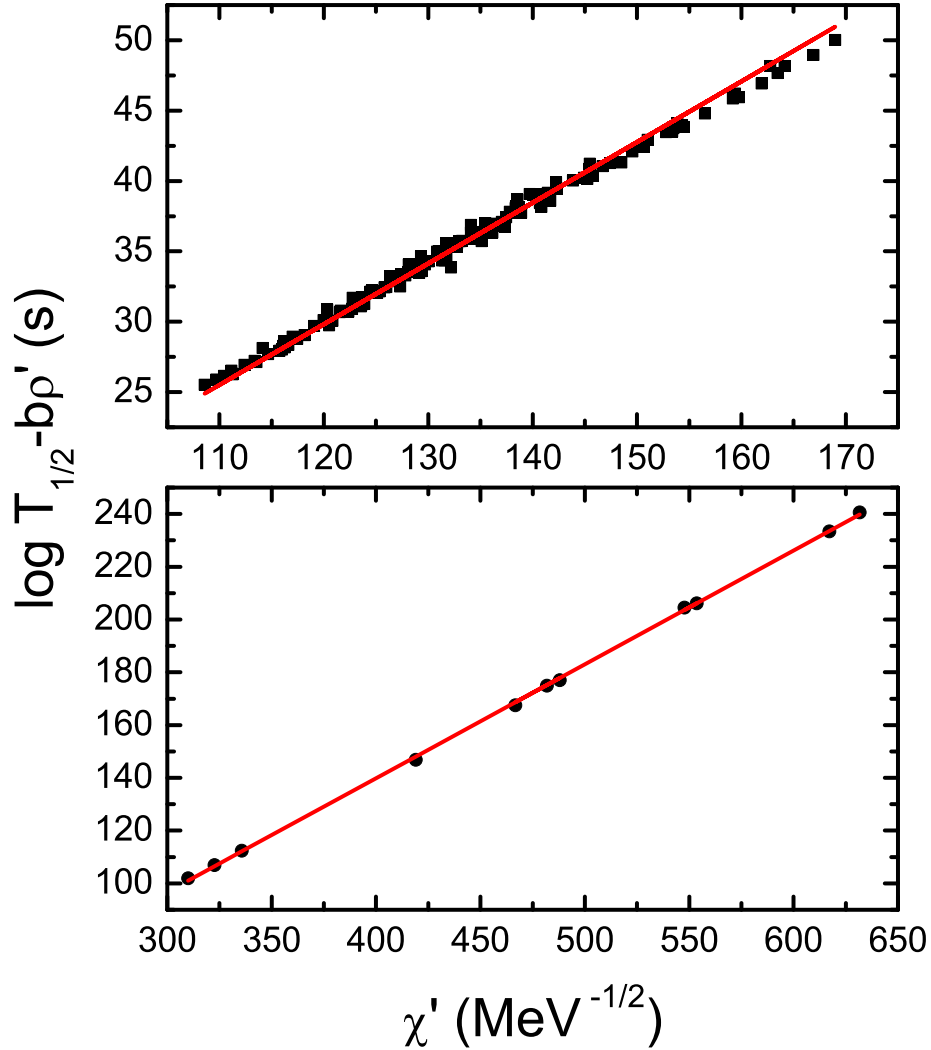}\\
\caption{(Color online) UDL description of $\alpha$ decays (upper
panel) and heavier cluster decays (lower panel) with the coefficient
$a$ fixed at its calculated value. The black points correspond to
experimental data with decay half-lives given in seconds. The lines
are given as $a\chi' + c$ with $c$ values from the lower part of
Table~\ref{table1}.}\label{fig4}
\end{figure}

The prediction power of the UDL (Eq.~(\ref{gn-2})) on radioactive
decay of medium and heavier nuclei has already been shown in our
previous Letter~\cite{qi08}. Essentially, only the coefficients $b$
and $c$ are free parameters not provided by the UDL. Without loss of
generalization, in that paper all coefficients of Eq.~(\ref{gn-2})
are determined by fitting experimental data. The inclusion of $a$ as
a free parameter takes into account the effect of higher-order terms
of the Coulomb penetrability. For example, the constants $a$, $b$
and $c$ corresponding to $\alpha$-decay were determined to be
0.4065, -0.4311 and -20.7889, respectively~\cite{qi08}. The standard
root mean square (rms) deviation between the UDL and experimental
$\alpha$-decay half-lives is $\sigma=0.3436$. For the rms deviation
we take the definition of Ref.~\cite{Poe06},
\begin{equation}
\sigma=\left\{\frac{1}{n-1}\sum_i^n\left[ \log (T^{\rm
Cal.}_i/T^{\rm Expt.}_i)\right]^2\right\}^{1/2},
\end{equation}
where $n$ is the number of decay events included in the fit and
$T^{\rm Expt.}$ and $T^{\rm Cal.}$ the experimental and calculated
decay half-lives, respectively. The fitted value for the coefficient
$a$ is close to the value calculated by its definition in
Eqs.~(\ref{gn-1}) and (\ref{gn-2}), namely $a =
e^2\pi\sqrt{2m}/(\hbar\ln10) = 0.4314$. Even if we fix the
coefficient $a$ at this value, the description power of the UDL is
still encouraging, as illustrated in the upper plot of
Fig.~\ref{fig4}. With only two free parameters of $b$ and $c$, the
UDL can reproduce experimental $\alpha$ decay half-lives with a rms
deviation of $\sigma=0.4606$.

In some cases the experimental half-lives are noticeably
underestimated by the UDL, with $T_{1/2}^{\rm Expt.}/T^{\rm
Cal.}>4$. These correspond to the $\alpha$ decays of $^{194,210}$Pb,
$^{208,210}$Po and $^{212}$Rn. This deviation may be related to the
fact that the formation amplitudes in these nuclei, due to the shell
closures of $N=126$ and $Z=82$, are significantly smaller than those
in the open shell region.

\begin{figure}
\includegraphics[scale=0.6]{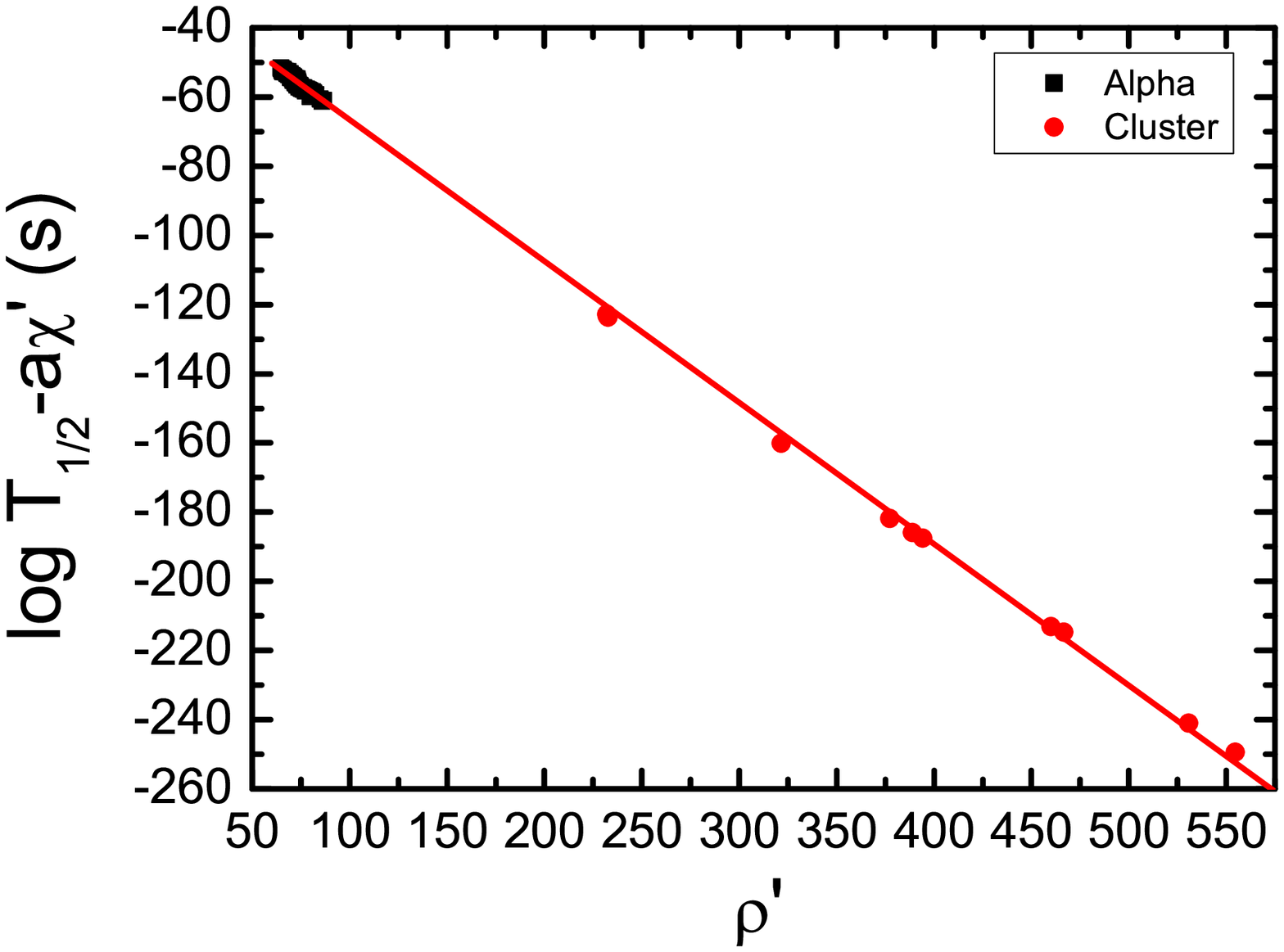}\\
\caption{(Color online) UDL description of both $\alpha$ and heavier
cluster decays as a function of $\rho'$. The lines are given as
$b\rho' + c$ with coefficients (set III) from the lower part of
Table~\ref{table1}.}\label{fig5}
\end{figure}

In fig.~\ref{fig4} we plotted calculations with the UDL on $\alpha$
and heavier cluster decay half-lives and comparisons with
experimental data. In these plots, the coefficient $a$ of the UDL is
taken as its calculated value while $b$ and $c$ are determined by
fitting to corresponding experiments. In the figure we plotted the
quantity $\log T_{1/2}-b\rho'$ as a function of $\chi'$. Similar
linear trend can be achieved if we plot experimental data as a
function of $\rho'$, as seen in Fig.~\ref{fig5} where all $\alpha$
and heavier cluster decays are considered. For all observed $\alpha$
and heavier cluster decays, the $\chi'$ and $\rho'$ values are in
the wide ranges of $105<\chi'<640$ and $60<\rho'<660$, respectively.
As a result, the functions $\log T_{1/2}-b\rho'$ and $\log
T_{1/2}-a\chi'$ we plotted in Figs.~\ref{fig4} and \ref{fig5} change
over 200 orders of magnitude. But the decay half-lives are in the
range of $-8<\log T_{1/2}<28$~(in seconds).

In Table~\ref{table1} the constants $a$, $b$ and $c$ that fit the
data sets of $\alpha$ as well as cluster decays are collected. The
fitted values with $a$ as a free and fixed parameter are shown in
the upper and lower part of the Table, respectively. In the table we
also give the corresponding rms deviations between experiments and
UDL calculations with these coefficient sets.

\begin{table}
\centering \caption{Upper: Coefficient sets of Eq.~(\ref{gn-2}) that
determined by fitting to experiments of $\alpha$ decays (I), cluster
decays (II) and both $\alpha$ and cluster decays (III),
respectively~\cite{qi08}, and the corresponding rms deviations;
Lower: same as the upper part but with coefficient $a$ fixed to its
calculated value of $a=0.4314$.\label{table1}}
\begin{ruledtabular}
\begin{tabular}{cccc}
& I($\alpha$) & II(cluster) & III($\alpha$+cluster)\\
\hline
a & ~~~0.4065 &~~~0.3671 & ~~~0.3949\\
b & ~~-0.4311 &~~-0.3296 & ~~-0.3693\\
c &-20.7889 &-26.2681 & -23.7615\\
$\sigma$ & ~~~0.3436 & ~~~0.6080 &~~~0.6107\\
\hline
a & ~~~0.4314 &~~~0.4314 & ~~~0.4314\\
b & ~~-0.4608 &~~-0.3921 & ~~-0.4087\\
c &-21.9453 &-32.7044 & -25.7725\\
$\sigma$ & ~~~0.4606 & ~~~0.7901 &~~~0.7631
\end{tabular}
\end{ruledtabular}
\end{table}

\section{Predictions and discussions}

Using the UDL it is straightforward to evaluate the half-lives of
all cluster emitters throughout the nuclear chart if reliable values
of the binding energies (i.e., of the cluster $Q$-values) can be
obtained. We do this by using the latest compilation of nuclear
masses~\cite{Audi03a}. With the $Q$-values thus obtained we
evaluated the decay half-lives of all isotopes included in that
compilation by applying the UDL. We will first show the case of the
decay of $\alpha$ particles and afterwards that of other relevant
clusters. For simplicity in what follows, only results calculated
with coefficients from the upper part of Table~\ref{table1} are
shown.

\begin{figure}
\includegraphics[scale=0.6]{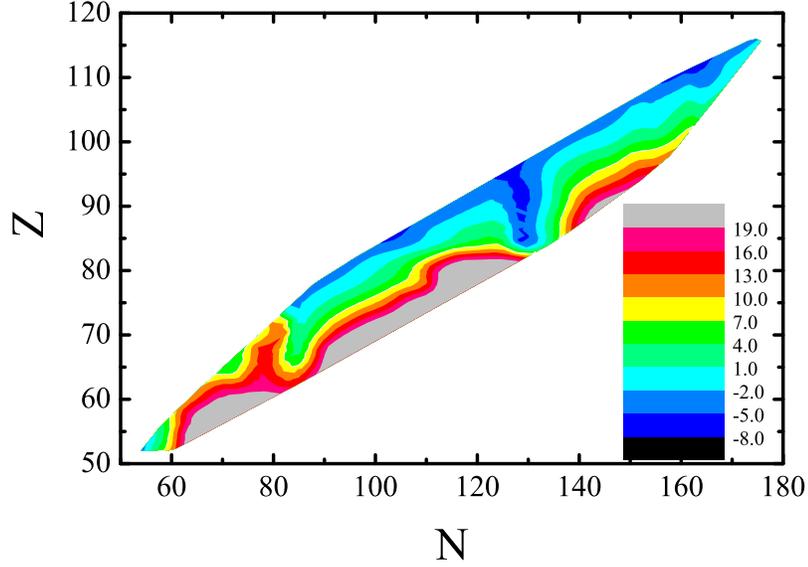}\\
\caption{(Color online) Prediction of the UDL (with the coefficient
set I) on the logarithms of half-lives (in seconds), $\log T_{1/2}$,
for the $\alpha$ decays of even-even nuclei.}\label{fig6}
\end{figure}

Since the half-lives of decaying nuclei which live a very short or
very large time can not be measured we will only consider even-even
$\alpha$ emitters with half-lives in the 30 orders of magnitude
range $-10\leq\log T_{1/2}\leq20$~(in second). The UDL predictions
of the corresponding half-lives are shown in  Fig.~\ref{fig6},
employing the coefficient set I from the upper part of
Table~\ref{table1}. Within the constraints that we imposed a total
number of 269 even-even $\alpha$ emitters have been found, which
have charge numbers $Z\geq52$. It is seen from the Figure that the
most favored $\alpha$ decays are from neutron-deficient nuclei
around the trans-lead and superheavy regions.

Since, as mentioned above, the emitters that we have used to
determine the coefficients of the UDL (Table~\ref{table1}) have
charge number $Z\geq78$, it would be interesting to probe the law
for nuclei with $Z$-values below that limit. We thus took the
extreme cases of decays from nuclei in the trans-tin region. The
$\alpha$ decay properties of nuclei in this region has been
intensively studied in recent years~\cite{Sili01,Janas05,Mazz02}. In
Table~\ref{table2} we compare experimental results on the $\alpha$
decays of Te, Xe and Ba isotopes and the predictions of the UDL. One
sees that in all cases the experimental values lie between the ones
calculated by using the parameters of the sets I and III in
Table~\ref{table1}, confirming the prediction power of the UDL.

\begin{table}
\centering \caption{Experimental and UDL calculated values (with the
coefficient sets I and III) of $\alpha$ decay half-lives (in
seconds) of even-even nuclei in the trans-tin region. Experimental
data are taken from Ref.~\cite{Audi03} except for the half-life of
$^{110}$Xe which is from Ref.~\cite{Janas05}.\label{table2}}
\begin{ruledtabular}
\begin{tabular}{ccccc}
Emitter& $Q_{\alpha}$(MeV) & $\log T_{1/2}^{{\rm Expt.}}$ & $\log T_{1/2}^{{\rm Cal.}}$(I) & $\log T_{1/2}^{{\rm Cal.}}$(III)\\
\hline
$^{106}$Te & 4.290 &~-4.155 &~-3.446  &   ~~-4.484    \\
$^{108}$Te & 3.445 &  ~0.6320 & ~0.9761 &  -0.1812 \\
$^{110}$Xe & 3.885 & -0.7850 & -0.3774 & ~~-1.441  \\
$^{112}$Xe & 3.330 & ~~2.477 &  ~~2.951 &  ~~~1.799    \\
$^{114}$Ba &3.534 &~~1.770 & ~~2.861    &   ~~~1.766\\
\end{tabular}
\end{ruledtabular}
\end{table}

\begin{figure}
\includegraphics[scale=0.6]{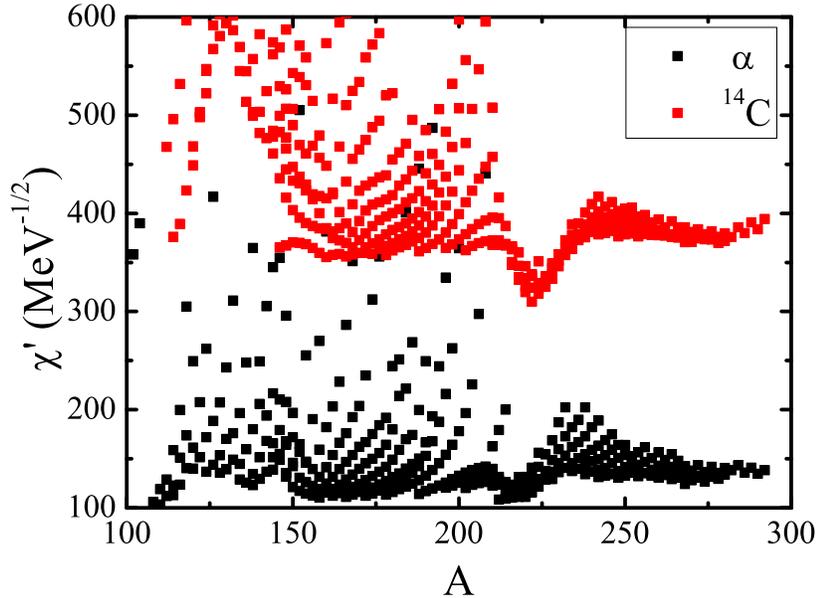}\\
\caption{(Color online) $\chi'$ values for $\alpha$ and $^{14}$C
radioactivities as a function of the mass numbers of mother
nuclei.\label{sch}}
\end{figure}

We will now apply the UDL to evaluate the emissions of heavy
clusters which are good candidates to be observed, namely
$^{12,14}$C, $^{16,18,20}$O, $^{20,22,24}$Ne, $^{24,26,28}$Mg and
$^{28,30,32,34}$Si. Observed cluster radioactivities exhibit much
longer partial half-lives than those of the corresponding $\alpha$
decays. This can be easily understood if we compare the $\chi'$
values of the heavier cluster and $\alpha$ radioactivities since the
logarithm of the half-life is proportional to $\chi'$. As a typical
example, in Fig.~\ref{sch} we plotted the $\chi'$ values of $\alpha$
and $^{14}$C decays as a function of the mass numbers of the mother
nuclei. The $\chi'$ values of heavier clusters are mostly much
higher than those of the corresponding $\alpha$ decays, indicating
that it is more difficult for the heavier clusters to penetrate
through the Coulomb barrier. Besides, from the figure one sees that
nuclei favoring cluster decays should mostly be located in the
trans-lead region.

\begin{figure}
\includegraphics[scale=0.6]{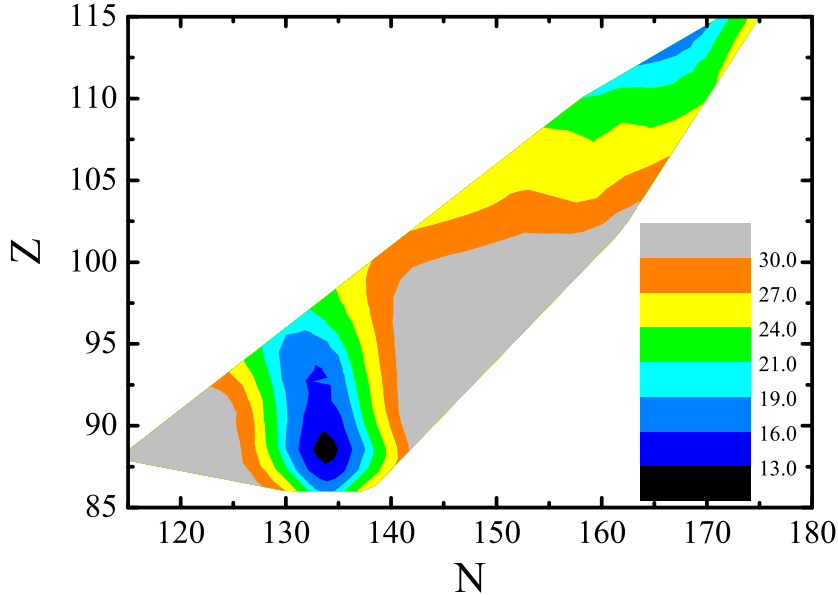}\\
\caption{(Color online) Same as Fig.~\ref{fig6} but for the $^{14}$C
cluster decay and with the coefficient set II.\label{c14}}
\end{figure}

In Fig.~\ref{c14} we show the predicted half-lives corresponding to
the most favored cluster radioactivity, namely $^{14}$C decay. Our
calculations show that nuclei like $^{220,222,224}$Ra,
$^{222,224}$Th and $^{226}$U can have partial decay half-lives
shorter than $10^{16}$~s, among which the $^{14}$C decays of
$^{222,224}$Ra have been observed~\cite{Rose84,Bonetti07}.

\begin{figure}
\includegraphics[scale=0.6]{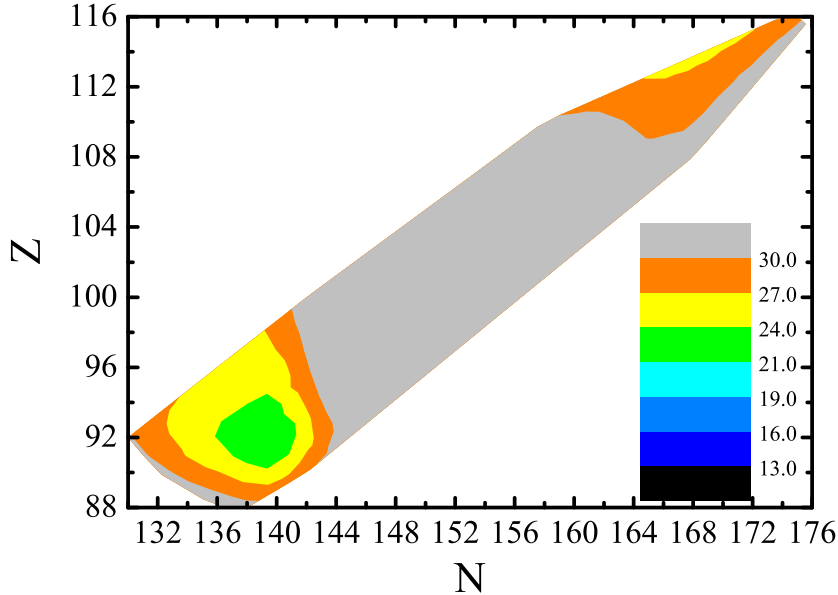}\\
\caption{(Color online) Same as Fig.~\ref{fig6} but for the
$^{24}$Ne decay.\label{ne24}}
\end{figure}

Our calculations also show that nuclei that most probably emit
clusters with non-equal proton and neutron numbers like $^{14}$C are
concentrated in the trans-lead region. This is consistent with the
expectation from the schematic picture of Fig.~\ref{sch}. For
heavier clusters the formation probability is even smaller and
therefore the corresponding decay probability is also smaller. As
another typical example, in Fig.~\ref{ne24} we plotted calculations
for the half-lives of the $^{24}$Ne radioactivity. One sees in this
Figure that the shortest half-lives correspond to mother nuclei
around Z=92 and N=138. In all cases this half-life is larger than
$10^{21}$ s, which is many orders of magnitude larger than the cases
corresponding to the decay of $^{14}$C analyzed above.

All heavier-cluster-decaying nuclei decay also by emitting $\alpha$
particles. In fact $\alpha$ decay is usually the overwhelming
dominant decay channel, as seen from Fig.~\ref{sch}. Therefore in
planning the detection of a probable cluster decay one has to
consider carefully the branching ratio of the $\alpha$-decay channel
relative to the corresponding cluster decay of interest, i.e.,
$b^{{\rm rel.}}=T^{\alpha}_{1/2}/T^{{\rm cluster}}_{1/2}$. The $\log
b^{{\rm rel}}$ values are negative, which can not be too small for
the heavier cluster decay to be detectable. We can evaluate these
branching ratios by using the UDL. To search for probable cluster
emitters we select particle decay channels for which neither the
half-lives are too large nor the branching ratios are too small. We
thus use the criteria $T_{1/2}<10^{30}$~s and $b^{{\rm
rel.}}>10^{-18}$ which is two orders of magnitude outside present
experimental limits. The corresponding calculations for the
emissions of $N_c\neq Z_c$ clusters are listed in
Table~\ref{table3}. To give an insight of the expected precision, we
present in Table \ref{table3a} comparisons between calculations and
experiments for the half-lives of the eleven observed heavier
cluster decay events of Fig.~\ref{fig4}.

\begin{table}
\centering \caption{Predictions of the UDL on probable emissions of
$N_c\neq Z_c$ clusters.\label{table3}}
\begin{ruledtabular}
\begin{tabular}{ccccc}
Emitter& Mode & $Q_{c}$(MeV) & $\log T_{1/2}^{{\rm cluster}}$ (s) & $-\log b^{{\rm rel.}}$\\
\hline
$^{220}$Rn  &$^{14}$C  &    28.539&    17.759&    15.573\\
$^{222}$Rn  &$^{14}$C  &    26.451&    22.313&    16.399\\
$^{220}$Ra  &$^{14}$C  &    31.038&    14.776&    16.142\\
$^{222}$Th  &$^{14}$C  &    31.653&    15.466&    17.776\\
$^{224}$Th  &$^{14}$C  &    32.930&    13.057&    12.657\\
$^{226}$Th  &$^{14}$C  &    30.547&    17.454&    13.747\\
$^{228}$Th  &$^{14}$C  &    28.222&    22.278&    14.082\\
$^{230}$Th  &$^{14}$C  &    26.060&    27.340&    14.605\\
$^{226}$U~  &$^{14}$C  &    32.969&    14.774&    14.993\\
$^{228}$U~  &$^{14}$C  &    30.525&    19.394&    16.308\\
$^{230}$U~  &$^{14}$C  &    28.339&    24.025&    17.331\\
$^{228}$Pu  &$^{14}$C  &    32.968&    16.572&    16.858\\
$^{226}$Th  &$^{18}$O  &    45.727&    18.235&    14.529\\
$^{228}$Th  &$^{18}$O  &    42.282&    23.933&    15.737\\
$^{230}$Th  &$^{18}$O  &    39.193&    29.674&    16.938\\
$^{228}$U~  &$^{18}$O  &    45.959&    20.083&    16.996\\
$^{226}$Ra  &$^{20}$O  &    40.817&    26.217&    15.189\\
$^{230}$Th  &$^{20}$O  &    41.795&    26.762&    14.026\\
$^{232}$U~  &$^{22}$Ne &    57.364&    26.532&    16.784\\
$^{232}$Pu  &$^{22}$Ne &    62.343&    21.941&    17.671\\
$^{228}$Th  &$^{24}$Ne &    57.414&    25.393&    17.197\\
$^{232}$Th  &$^{24}$Ne &    54.497&    29.916&    11.951\\
$^{230}$U~  &$^{24}$Ne &    61.352&    22.171&    15.477\\
$^{234}$U~  &$^{24}$Ne &    58.826&    25.727&    12.542\\
$^{234}$Pu  &$^{24}$Ne &    62.254&    23.382&    17.336\\
$^{232}$U~  &$^{26}$Mg &    71.771&    27.481&    17.732\\
$^{232}$Pu  &$^{26}$Mg &    78.366&    21.852&    17.583\\
$^{234}$Pu  &$^{26}$Mg &    78.313&    21.786&    15.739\\
$^{232}$U~  &$^{28}$Mg &    74.320&    25.201&    15.453\\
$^{234}$Pu  &$^{28}$Mg &    79.154&    21.807&    15.760\\
$^{238}$Pu  &$^{28}$Mg &    75.912&    25.800&    16.154\\
$^{238}$Cm  &$^{28}$Mg &    80.368&    23.023&    17.547\\
$^{238}$Cm  &$^{30}$Si &    95.577&    22.601&    17.125\\
$^{236}$Pu  &$^{32}$Si &    91.674&    24.941&    16.741\\
$^{238}$Cm  &$^{32}$Si &    97.262&    21.513&    16.037\\
$^{240}$Cm  &$^{32}$Si &    97.555&    21.020&    14.559\\
$^{238}$Pu  &$^{34}$Si &    90.812&    26.753&    17.106\\
$^{240}$Pu  &$^{34}$Si &    91.029&    26.322&    14.728\\
$^{240}$Cm  &$^{34}$Si &    95.468&    24.290&    17.829\\
\end{tabular}
\end{ruledtabular}
\end{table}

\begin{table}
\centering \caption{Experimental and UDL calculated values (with the
coefficient set II of cluster decay half-lives (in seconds). The
experimental values are from Ref.~\cite{Bonetti07}.\label{table3a}}
\begin{ruledtabular}
\begin{tabular}{ccccc}
Emitter& Mode & $Q_{c}$(MeV) & $\log T_{1/2}^{{\rm Expt.}}$ &$\log T_{1/2}({\rm II})$\\
\hline
$^{222}$Ra  &$^{14}$C  &    33.05  &11.01 &  11.07\\
$^{224}$Ra  &$^{14}$C  &    30.54  &15.86 &  15.59\\
$^{226}$Ra  &$^{14}$C  &    28.20  &21.24 &  20.33\\
$^{228}$Th  &$^{20}$O  &    44.72  &20.72 &  21.59\\
$^{230}$U~  &$^{22}$Ne &    61.39  &19.22 &  20.73\\
$^{230}$Th  &$^{24}$Ne &    57.76  &24.61 &  24.74\\
$^{232}$U~  &$^{24}$Ne &    62.31  &20.40 &  20.68\\
$^{234}$U~  &$^{28}$Mg &    74.11  &25.75 &  25.36\\
$^{236}$Pu  &$^{28}$Mg &    79.67  &21.52 &  21.02\\
$^{238}$Pu  &$^{32}$Si &    91.19  &25.27 &  25.39\\
$^{242}$Cm  &$^{34}$Si &    96.51  &23.15 &  22.87\\
\end{tabular}
\end{ruledtabular}
\end{table}

The $\alpha$-decay mode dominates the decays of all heavier-cluster
emitters we listed in Table~\ref{table3}. In most cases the
branching ratio between $\alpha$ decay and all other decay channels
(including $\beta$ decay) is $b^{\alpha}\simeq100\%$~\cite{Audi03}.
But there are exceptions, in particular the nuclei $^{232}$Pu,
$^{234}$Pu and $^{238}$Cm which have the $\alpha$-decay branching
ratios of $b^{\alpha}=11\%$, $b^{\alpha}=6\%$ and $b^{\alpha}
<10\%$, respectively~\cite{Audi03}.

\begin{figure}
\includegraphics[scale=0.6]{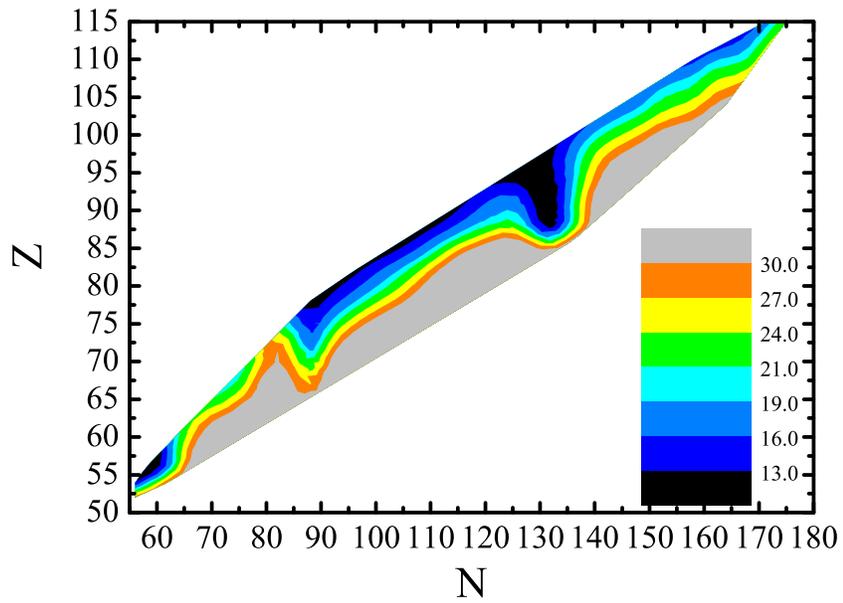}\\
\caption{(Color online) Same as Fig.~\ref{fig6} but for the $^{12}$C
decay.\label{c12}}
\end{figure}

We will now analyze the more rare case of radioactive decays of
$N_c=Z_c$ clusters heavier than the $\alpha$ particle. Intense
studies have been made in the prediction and searching for the
emissions of $N_c=Z_c$
clusters~\cite{Sili01,Mazz02,Ogan94,Gug95,Gug97,Poe93,Kumar95}.
Experiments have not pinned down the observation of these clusters
yet, although efforts have been made, particularly looking for the
probable emission of $^{12}$C~\cite{Gug97}. We have therefore apply
the UDL to investigate regions in the nuclear chart where such
cluster would be likely to be formed and emitted. The half-lives of
$^{12}$C decays thus calculated are plotted in Fig.~\ref{c12}. The
emissions of other $N_c=Z_c$ clusters like $^{16}$O show similar
patterns. It is seen from the figure that $N_c=Z_c$ cluster emitters
form two islands, decaying into daughter nuclei around $^{100}$Sn
and $^{208}$Pb. This is consistent with theoretical calculations
using the fission model~\cite{Grei89,Poe93}.

\begin{figure}
\includegraphics[scale=0.6]{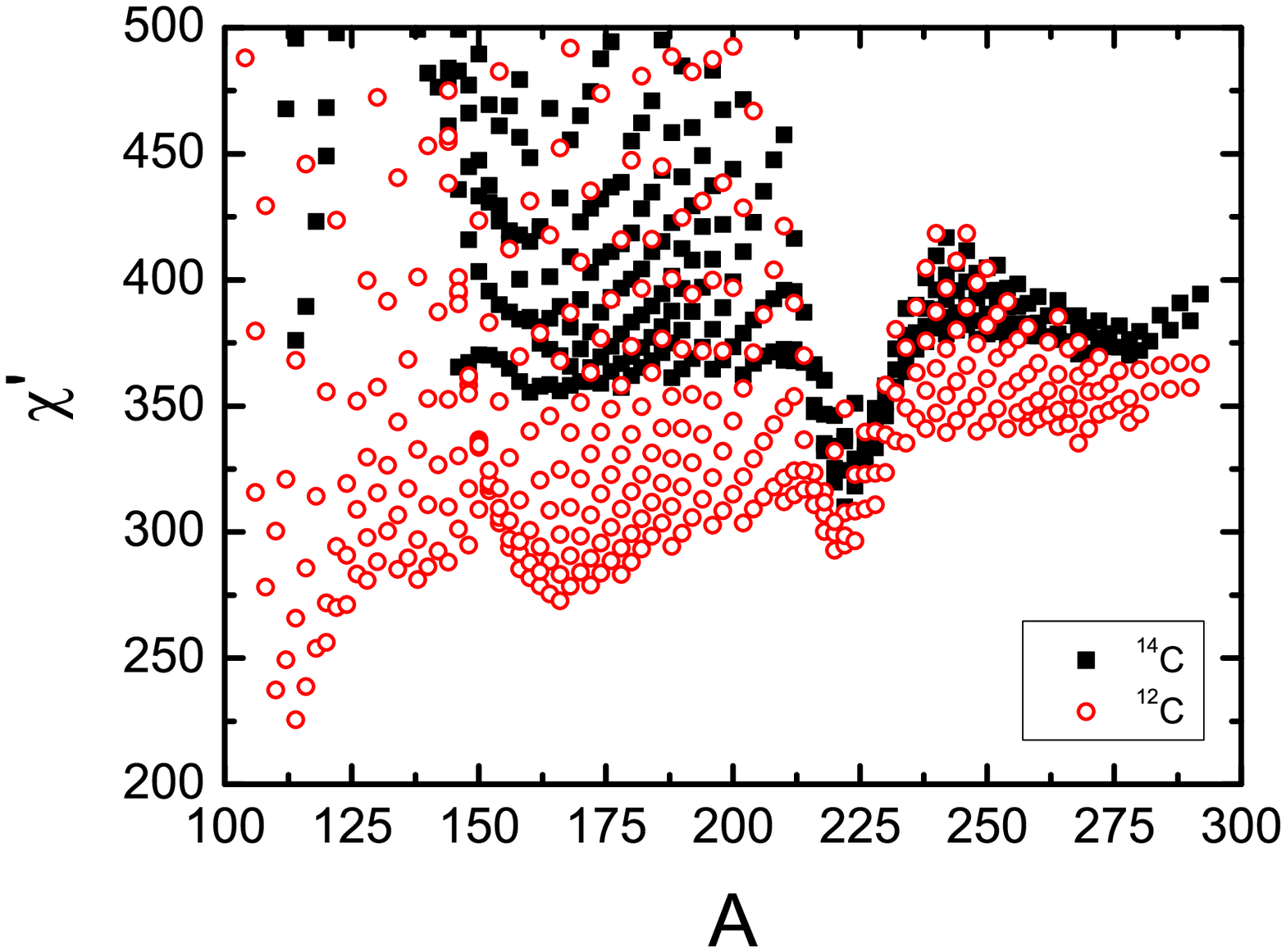}\\
\caption{(Color online) $\chi'$ values for $^{12}$C and $^{14}$C
radioactivities as a function of the mass numbers of mother
nuclei.\label{c1214}}
\end{figure}

A first glance at Fig.~\ref{c12} may suggest that the emissions of
$N_c=Z_c$ clusters like $^{12}$C should be more favored than those
of other $N_c\neq Z_c$ isotopes since the former particle is usually
more tightly bound. Such a picture is also expected if we compare
the $\chi'$ values for the radioactive decays of other isotopes. A
typical example is given in Fig.~\ref{c1214} where we plotted the
$\chi'$ values of the $^{12}$C and $^{14}$C radioactivities. It is
seen that the $\chi'$ values of the $^{12}$C radioactivity are
mostly smaller that those of $^{14}$C, indicating that it should be
much easier for the $^{12}$C particle to penetrate through the
Coulomb barrier, especially in nuclei close to the proton drip line.
However, the probability of the decay of $N_c=Z_c$ clusters become
small if we take into account the fact that the likely emitters are
mostly close to the proton drip line and are dominated by the decay
mode of $\beta^+$. With the same selection criteria discussed above,
our predictions on probable emissions of $N_c=Z_c$ clusters are
listed in Table~\ref{table4}. Since in all cases the decay by the
emission of an $\alpha$ particle is much more likely than the
corresponding decay by the emission of heavier clusters, in
Table~\ref{table4} we only show emitters that are known to decay
$\alpha$~\cite{Audi03}. It is seen from the Table that the mostly
likely $N_c=Z_c$ cluster emitter is the nucleus $^{114}$Ba.

\begin{table}
\centering \caption{Predictions of the UDL on probable emissions of
$N_c=Z_c$ clusters. The $\alpha$ decay branching ratios,
$b^{\alpha}$, of the cluster emitters are taken from
Ref.~\cite{Audi03}.\label{table4}}
\begin{ruledtabular}
\begin{tabular}{cccccc}
Emitter& Mode & $Q_{c}$(MeV) & $\log T_{1/2}^{{\rm cluster}}$(s) & $-\log b^{{\rm rel.}}$ & $b^{\alpha}$(\%)\\
\hline
$^{110}$Xe  &$^{12}$C  &    15.726&    12.863&    13.241   &64\\
$^{112}$Xe  &$^{12}$C  &    14.283&    17.099&    14.148  &0.9\\
$^{114}$Ba  &$^{12}$C  &    18.984&    ~7.199&    ~4.338  &0.9\\
$^{154}$Dy  &$^{12}$C  &    15.557&    28.193&    14.432  &100\\
$^{158}$Yb  &$^{12}$C  &    20.078&    19.374&    12.888  &2.1E-3\\
$^{160}$Hf  &$^{12}$C  &    21.922&    17.015&    13.640  &0.7\\
$^{162}$Hf  &$^{12}$C  &    20.144&    21.415&    15.362  &8E-3\\
$^{162}$W   &$^{12}$C  &    23.831&    14.813&    14.100  &45.2\\
$^{166}$W   &$^{12}$C  &    20.720&    22.080&    17.505  &3.5E-2\\
$^{166}$Os  &$^{12}$C  &    24.495&    15.339&    15.644  &72\\
$^{168}$Os  &$^{12}$C  &    23.274&    17.959&    17.000  &49\\
$^{166}$Pt  &$^{12}$C  &    27.941&    10.616&    14.039  &100\\
$^{168}$Pt  &$^{12}$C  &    26.815&    12.619&    15.159  &100\\
$^{170}$Pt  &$^{12}$C  &    25.799&    14.537&    16.133  &8.6\\
$^{172}$Pt  &$^{12}$C  &    24.836&    16.463&    17.222  &72\\
$^{172}$Hg  &$^{12}$C  &    28.275&    11.680&    15.146  &100\\
$^{174}$Hg  &$^{12}$C  &    27.355&    13.311&    15.905  &100\\
$^{176}$Hg  &$^{12}$C  &    26.454&    14.993&    16.504  &90\\
$^{180}$Hg  &$^{12}$C  &    24.645&    18.666&    17.879  &48\\
$^{178}$Pb  &$^{12}$C  &    29.006&    12.013&    15.597  &100\\
$^{180}$Pb  &$^{12}$C  &    28.052&    13.691&    16.164  &100\\
$^{184}$Pb  &$^{12}$C  &    26.193&    17.231&    17.600  &80\\
$^{202}$Ra  &$^{12}$C  &    29.630&    15.569&    17.955  &100\\
$^{218}$Ra  &$^{12}$C  &    30.436&    13.497&    17.715  &100\\
$^{220}$Ra  &$^{12}$C  &    32.021&    10.662&    12.027  &100\\
$^{222}$Ra  &$^{12}$C  &    29.049&    15.957&    14.029  &100\\
$^{224}$Ra  &$^{12}$C  &    26.375&    21.476&    15.609  &100\\
$^{226}$Ra  &$^{12}$C  &    23.850&    27.524&    16.496  &100\\
$^{220}$Th  &$^{12}$C  &    32.139&    12.227&    16.852  &100\\
$^{222}$Th  &$^{12}$C  &    33.156&    10.455&    12.765  &100\\
$^{224}$Th  &$^{12}$C  &    30.366&    15.248&    14.848  &100\\
$^{226}$Th  &$^{12}$C  &    27.667&    20.569&    16.863  &100\\
$^{222}$U~  &$^{12}$C  &    33.897&    10.968&    16.320  &100\\
$^{224}$U~  &$^{12}$C  &    34.373&    10.132&    13.178  &100\\
$^{226}$U~  &$^{12}$C  &    31.649&    14.651&    14.869  &100\\
$^{228}$U~  &$^{12}$C  &    28.969&    19.714&    16.627  &$>95$\\
$^{228}$Pu  &$^{12}$C  &    32.797&    14.327&    14.613  &100\\
$^{112}$Xe  &$^{16}$O  &    21.000&    20.519&    17.568  &0.9\\
$^{114}$Ba  &$^{16}$O  &    26.422&    11.477&    ~8.616  &0.9\\
$^{162}$Hf  &$^{16}$O  &    31.657&    21.563&    15.510  &8E-3\\
$^{166}$Os  &$^{16}$O  &    37.132&    16.535&    16.839  &72\\
$^{168}$Pt  &$^{16}$O  &    40.005&    14.214&    16.754  &100\\
$^{172}$Hg  &$^{16}$O  &    41.502&    14.053&    17.518  &100\\
$^{224}$Th  &$^{16}$O  &    46.482&    15.321&    14.921  &100\\
$^{226}$Th  &$^{16}$O  &    42.662&    21.196&    17.489  &100\\
$^{226}$U~  &$^{16}$O  &    48.019&    15.152&    15.371  &100\\
$^{228}$U~  &$^{16}$O  &    44.331&    20.657&    17.570  &$>95$\\
$^{228}$Pu  &$^{16}$O  &    49.485&    15.095&    15.381  &100\\
\end{tabular}
\end{ruledtabular}
\end{table}

\section{Summary and Conclusions}

Starting from the exact expression for the half-life of cluster
decaying nuclei (Eq.~(\ref{life})) we found that this expression is
dependent upon a quantity called $\cos^2\beta$ which for medium and
heavier nuclei is small (Eq.~(\ref{cosb})). For the case of $l=0$
(monopole) transitions we expanded the exact expression  to the
lowest order in $\cos^2\beta$ and used the property that the
half-life does not depend upon the matching radius $R$
(Eq.~(\ref{foramp})). We thus found that the logarithm of the
half-life is linearly dependent upon two parameters, called $\chi'$
and $\rho'$, which depend only upon the $Q$-value of the outgoing
cluster and of the charges and masses of the particles involved in
the decay (Eq.~(\ref{chirhop})). The resulting linear expression
(Eq.~(\ref{gn-2})) is found to be a generalization of the
Geiger-Nuttall law and we call it the universal decay law (UDL). The
UDL is valid for all $l=0$ transitions. This monopole linear
equation contains three constants, called $a$, $b$ and $c$. We
fitted the experimental half-lives of ground-state to ground-state
$\alpha$-decay and heavier cluster decay processes in even-even
nuclei to obtain the values of the constants given in
Table~\ref{table1}. We found that the UDL predicts with great
precision the half-lives of radioactive decays, both $\alpha$- and
cluster-decays, and for all isotopic series, as expected since the
original exact expression for the half-life is valid in general.
This law may also help in the ongoing search of new cluster decay
modes from superheavy nuclei.

Using the UDL we have evaluated the decay half-lives of various
cluster emitters throughout the nuclear chart with reliable values
of binding energies as input. It is found that the $\alpha$ decay is
favored in neutron-deficient nuclei around the trans-lead and
superheavy regions. The decays of heavier clusters with non-equal
proton and neutron numbers are mostly located in the trans-lead
region. The probability of the decay of clusters with equal numbers
of protons and neutrons is small since the likely emitters are
mostly close to the proton drip line and are dominated by the decay
mode of $\beta^+$.

An important conclusion from the UDL is that the cluster formation
amplitude $F_c(R)$ is exponentially dependent upon the variable
$\rho'$. The implication of this linear trend on nuclear structure
effects may deserve further investigations in the future.

\section*{Acknowledgments}

This work has been supported by the Chinese Major State Basic
Research Development Program under Grant 2007CB815000; the National
Natural Science Foundation of China under Grant Nos. 10525520,
10735010 and 10875172 and the Swedish Science Research Council (VR).

\end{document}